\RequirePackage{fix-cm}
\documentclass[smallextended]{svjour3}       
\smartqed  
\usepackage{graphicx}
\usepackage{amsmath}
\usepackage{txfonts}
\usepackage{natbib}

\newcommand{\com}[1]{\textnormal{#1}}
\journalname{General Relativity and Gravitation}
\begin{document}

\title{Cosmic structures from a mathematical perspective \\ 1. Dark matter halo mass density profiles 
}


\author{Jenny Wagner         
}


\institute{J. Wagner (ORCID: 0000-0002-4999-3838) \at Universit\"{a}t Heidelberg, Zentrum f\"{u}r Astronomie, Astron. Rechen-Institut, \\ M\"{o}nchhofstr. 12--14, 69120 Heidelberg, Germany \\
               \email{j.wagner@uni-heidelberg.de}
}

\date{Received: date / Accepted: date}

\maketitle
\begin{abstract}
The shapes of individual self-gravitating structures of an ensemble of identical, collisionless particles have remained elusive for decades. 
In particular, a reason why mass density profiles like the Navarro-Frenk-White or the Einasto profile are good fits to simulation- and observation-based dark matter halos has not been found.
Given the class of three dimensional, spherically symmetric power-law probability density distributions to locate individual particles in the ensemble mentioned above, we derive the constraining equation for the power-law index for the most and least likely joint ensemble configurations. 
\com{We find that any dark matter halo can be partitioned into three regions: 
a core, an intermediate part, and an outskirts part up to boundary radius $r_\mathrm{max}$. 
The power-law index of the core is determined by the mean radius of the particle distribution within the core.
}
The \com{intermediate} region becomes isothermal in the limit of infinitely many particles.
The slope of the mass density profile far from the centre is determined by the \com{choice of $r_\mathrm{max}$ with respect to the outmost halo particle}, such that two typical limiting cases arise that explain the $r^{-3}$-slope for galaxy-cluster outskirts and the $r^{-4}$-slope for galactic outskirts.
Hence, \com{we succeed in deriving the mass density profiles of common fitting functions from a general viewpoint.} 
These results also allow to find a simple explanation for the cusp-core-problem and to separate the halo description from its dynamics.

\end{abstract}

\section{Introduction}
\label{sec:introduction}

During the past decades, large-scale N-body simulations have successfully reconstructed cosmic structure formation with increasing resolution and complexity, as observations corroborate, see e.g.\@ \cite{bib:Bullock}, \cite{bib:DelPop}, \cite{bib:Frenk}, or \cite{bib:Kuhlen} for summarising overviews.
Complementary efforts to search for a closed-form description have arrived at a hydrodynamical theory that explains cosmic structure evolution up to the non-linear regime, see \cite{bib:Schaefer} for a recent overview. 
A new approach to describe cosmic structure formation based on a kinetic field theory can derive an analytic, parameter-free equation for the non-linear cosmic power spectrum, see \cite{bib:Bartelmann_KFT} for an introduction. 

All of these approaches describe cosmic structures as evolving from initial overdensity seeds in the power spectrum that are propagated forward in cosmic time. 
To yield the non-linear agglomerations we observe today, these approaches thus require an initial configuration of overdensity seeds in phase-space together with a dynamical model of its evolution over cosmic time. 
As a result, they obtain the statistical properties of mass density perturbations of the observable universe as a whole based on the evolution of the phase-space-volume of the initial overdensity seeds.
Subsequent attempts to derive the shape of individual mass density profiles of simulated and observed galaxies or galaxy clusters from these approaches have not been successful yet.
A potential reason that the mass density profiles have not been explained by one of these approaches so far is the missing physical link between fluctuations in the power spectrum of the mass overdensity and the definition of an individual cosmic structure, like a (dark matter) halo. 

Several heuristic fitting functions have been developed that capture the shape of individual dark matter halos, among others the Navarro-Frenk-White profile \cite{bib:NFW1} and the Einasto profile \cite{bib:Einasto}, and the shape of individual luminous matter distributions, like the de-Vaucouleurs profile \cite{bib:Vaucouleurs} or the Jaffe profile \cite{bib:Jaffe}.
Without a deeper understanding how these profiles can be derived from more fundamental principles, it still remains an open question why these formulae are good fits to the simulated and observed mass density distributions.

To further investigate this question, we develop a statistical approach to derive the shape of dark matter halo density profiles from the framework of probability theory.
The approach is based on an ensemble of independent, identical dark matter particles that are spatially spread according to generic, spherically symmetric power-law probability density distributions which are typical for scale-free interactions like gravitation. 
Although we only consider ensembles of dark matter particles, the approach is generic and can be applied to any ensemble of particles that shares the same prerequisites.

Contrary to the approaches mentioned above, we split cosmic structure evolution into the description of structures at one instant in cosmic time and a second part describing their dynamical evolution. 
In this work, we focus on the former part and leave out the latter. 
This is feasible because we do not track the change of initial structures across cosmic time to model a current one. 
Instead, at one instant in cosmic time, we define a structure as an agglomeration of particles on a predefined length scale of interest.
This approach is supported by the simulations of \cite{bib:Moore}, showing that the mass density profiles of dark matter halos at one time seem to be independent of their merger history.
\com{\cite{bib:Hjorth} arrive at similar results, when setting up their derivation of dark matter halo mass density profiles. They find that approaches to explain dark matter halo mass density profiles based on statistical mechanics, including their own, deal with final equilibrium states of the halo mass densities. Hence, the resulting dark matter halo mass density profiles, like their ``DARKexp'' final equilibrium state, are independent of any dynamical relaxation process.}
 
Defining a (cosmic) structure as a particle ensemble at a specific scale enables us to set up a general characterisation of a dark matter halo that is solely subject to gravitational forces acting on this scale. 
Our effective approach is thus independent of the processes happening on smaller scales and of the dynamics on any scale leading to the state under consideration\com{,  similar to \cite{bib:Hjorth}, but without the need to introduce an energy-state-space or a phase-space.}
We find that the parametric fitting functions mentioned above can be derived from this general characterisation as combinations of limiting cases for the halo core, the isothermal middle part, and the boundary regions, taking into account the finiteness of the observation or simulation to be described, meaning their finite resolution, finite volume, and finite amount of dark matter particles.

The work is organised as follows: 
In Section~\ref{sec:profiles}, we briefly summarise the parametric mass density profiles for the most commonly used dark matter halo profiles of galaxies and galaxy clusters. 
Given a small set of assumptions as prerequisites, detailed in Section~\ref{sec:derivation}, we set up a maximum-likelihood approach to derive the power-law index for the ensemble of dark matter particles. 
In Section~\ref{sec:approximations}, limiting approximations of the general case set up in Section~\ref{sec:derivation} are considered.
We derive the power-law index for \com{the core}, \com{the isothermal, intermediate region}, and describe the boundary regions for galaxy-scale and galaxy-cluster-scale dark matter halos.
In Section~\ref{sec:transfer}, we set up a mass density profile from the particle ensemble, such that the mathematical results derived in the previous sections can be assigned a physically meaningful interpretation.
In this way, we can explain the shape of the most commonly employed halo mass density profiles of Section~\ref{sec:profiles}.
We conclude in Section~\ref{sec:conclusion} with a summary of the approach, its current explanatory power, and an outlook to the next parts of this series.

\section{Halo mass density profiles from a physical viewpoint}
\label{sec:profiles}

Our characterisation of cosmic structures decisively depends on the definition thereof. 
The most important property of a structure is that its constituents are gravitationally bound to each other. 
The structure should be of finite extent and separable from similar structures in its vicinity, unless structures merge. 
Furthermore, a structure requires a characteristic (length) scale.
Mathematically, a characteristic (length) scale serves as a scaling (radius) to obtain dimensionless variables.
Physically, the scaling (radius) is important to mark characteristic scales of potentially occurring phase transitions.

Tracing all individual constituents within cosmic structures and taking into account their mass to set up a mass density distribution of these structures is computationally infeasible for luminous matter and principally impossible for dark matter.
Therefore, we describe the mass density distribution of gravitationally bound, separable structures of finite extent by a continuous mass density $\rho(\boldsymbol{r})$, averaging the distribution of particles over scales that are larger than their mutual mean free path length and that are smaller than the extent of the structure, see \cite{bib:Binney} for details.

We briefly discuss the most common mass density profiles for galaxy- and galaxy-cluster-sized dark matter halos. 
An extended list can be found e.g.\@ in \cite{bib:Coe} and \cite{bib:Wagner3}.
All models are designed to characterise the mass density profile of a dark matter halo with a small number of parameters that have a physical interpretation.
We restrict our analysis to spherically symmetric mass density profiles, i.e.\@ we neglect the perturbations by interactions with the environment. 
Small perturbations are usually modelled by introducing elliptical isocontours, larger deviations from spherical symmetry often require to use more complex mass density profiles. 
The assumption of spherical symmetry is a good approximation to the inner parts of galaxy clusters and heavy, elliptical galaxies. 
Making this simplification, we can derive the most-likely parameter values analytically as a proof of concept.
These derivations can be extended to cases with less symmetry in a future work.

\subsection{Common mass density profiles for dark matter halos}
\label{sec:density_profiles}

\subsubsection{Navarro-Frenk-White profile}
\label{sec:NFW} 

The three-dimensional Navarro-Frenk-White (NFW) mass density profile, as introduced in \cite{bib:NFW1} and \cite{bib:NFW2}, reads
\begin{equation}
\rho_\mathrm{NFW}(r) = \dfrac{\rho_\mathrm{s}}{r/r_\mathrm{s} \left( 1 + r/r_\mathrm{s} \right)^2} \;,
\label{eq:NFW_rho}
\end{equation}
in which $r$ denotes the radius, $r_\mathrm{s}$ a scale radius. 
$\rho_\mathrm{s}$ is the scale density. 
It is four times the mass density at $r_\mathrm{s}$, i.e.~$\rho_\mathrm{s} = 4 \rho_\mathrm{NFW}(r_\mathrm{s})$.
As stated in \cite{bib:Frenk} and references therein, the physical origins of this profile have not yet been fully understood. 
Usually, $r_\mathrm{s}$ is defined as the radius for which the logarithmic slope of the mass density profile is -2. 

For a general limiting radius $r_\mathrm{max}$, the NFW-halo mass is given as
\begin{equation}
M(r_\mathrm{max}) = 4\pi \int \limits_0^{r_\mathrm{max}} \rho_\mathrm{NFW}(r) r^2 \mathrm{d} r = 4\pi \rho_\mathrm{s} r_\mathrm{s}^3 \left( \ln \left( 1 + \tfrac{r_\mathrm{max}}{r_\mathrm{s}} \right) - \tfrac{r_\mathrm{max}}{r_\mathrm{s} + r_\mathrm{max}} \right) \;.
\label{eq:NFWmass}
\end{equation}
Note that $M(0)=0$, despite the singular core of the mass density profile at $r=0$. 

Usually, Equation~\ref{eq:NFW_rho} is integrated to $r_\mathrm{max}=r_{200}$. This radius is determined by assuming that the mass of the dark matter halo up to $r_{200}$ equals the mass of a sphere with the same radius but a constant density which equals 200 times the critical density of the homogeneous cosmic background density, $\rho_{200}$.
This radius is chosen as an approximation to the radius at which the halo is in virial equilibrium, called $r_\mathrm{vir}$.

Instead of characterising the NFW profile by $\rho_\mathrm{s}$ and $r_\mathrm{s}$, it is common to use its mass given by Equation~\ref{eq:NFWmass} and the concentration parameter $c$
\begin{equation}
c  \equiv \dfrac{r_\mathrm{vir}}{r_\mathrm{s}} \approx \dfrac{r_{200}}{r_\mathrm{s}} \;.
\label{eq:NFW_c}
\end{equation}
With the definition of $c$ and $\rho_{200}$, Equation~\ref{eq:NFWmass} for $r_\mathrm{max}=r_{200} \approx r_\mathrm{vir}$ becomes
\begin{equation}
M(r_{200}) = \tfrac{4\pi}{3} \rho_{200} r_{200}^ 3 \approx 4\pi \rho_\mathrm{s} r_\mathrm{s}^3 \left( \ln \left( 1 + c \right) - \tfrac{c}{1+c} \right)\;.
\label{eq:NFWmassc}
\end{equation}
Fitting the NFW profile to equilibrium mass densities in simulations, \cite{bib:NFW2} found that it can describe these mass density profiles well over two decades in radius independent of the halo mass, the initial power spectrum, and the values of the cosmological parameters. 
Depending on the resolution of the data set, other mass density profiles may yield better fits.
Yet, the NFW profile is still one of the most commonly employed models of sufficient fitting quality for a lot of applications.

%

\subsubsection{Moore profile}
\label{sec:Moore} 

Another dark matter halo profile commonly discussed in the literature, is the Moore profile (MP, for short).
It is defined in \cite{bib:Moore} by
\begin{equation}
\rho_\mathrm{M}(r) = \dfrac{\rho_\mathrm{s}}{\left(r/r_\mathrm{s}\right)^{3/2} \left( 1 + (r/r_\mathrm{s})^{3/2} \right)} \;.
\label{eq:Moore_rho}
\end{equation}
As for Equation~\ref{eq:NFW_rho}, $r_\mathrm{s}$ denotes a scale radius, usually chosen as mentioned above.
While its behaviour for large radii is the same as for the NFW profile, it was defined to be a modification of the NFW profile with a steeper slope in the central part of the halo.

\subsubsection{Einasto profile}
\label{sec:Einasto} 

A third common three-dimensional dark matter halo profile form is given by the Einasto profile  (EP), as defined in \cite{bib:Einasto}. 
The profile has the form
\begin{equation}
\ln\left( \dfrac{\rho_\mathrm{E}(r)}{\rho_\mathrm{s}} \right) = -\dfrac{2}{\beta}\left( \left(\dfrac{r}{r_\mathrm{s}}\right)^{\beta} - 1  \right) \;,
\label{eq:Einasto_rho}
\end{equation}
in which the scaling radius $r_\mathrm{s}$ is usually chosen as mentioned above.
The additional third parameter of the profile, $\beta$, is called the shape parameter. 
As $\beta$ allows for a change in shape over the radius, which is not possible in the NFW or Moore profile, the Einasto profile has been found to better fit highly resolved N-body simulations of dark matter halos, \cite{bib:Merritt1}, \cite{bib:Merritt2}, \cite{bib:Gao}, \cite{bib:Hayashi}, \cite{bib:NFW3}.

\subsubsection{Pseudo-isothermal elliptical mass profile}
\label{sec:PIEMD} 

The fourth commonly used dark matter halo profile is the pseudo-isothermal elliptical mass distribution (PIEMD).  
It is introduced in \cite{bib:Kassiola} as
\begin{equation}
\rho_\mathrm{PIEMD}(r) = \dfrac{\rho_\mathrm{s}}{\left( 1 + (r/r_\mathrm{s})^2 \right) \left( 1 + (r/r_\mathrm{c})^2 \right)^{n}}
\label{eq:PIEMD_rho}
\end{equation}
with a scaling density $\rho_\mathrm{s}$ and two scaling radii, $r_\mathrm{s}$ and $r_\mathrm{c}$, denoting the scale radius of the halo core and of a cut-off radius to truncate the halo mass density, respectively.
In \cite{bib:Kassiola}, the radius $r$ is assumed to constrain an elliptical isocontour, however, we do not use the ellipticity here and assume spherically symmetric isocontours for the halo mass density. 
The exponent of the second term in the denominator is usually set to $n=1$.
The behaviour in the centre of the halo depends on the size of $r_\mathrm{s}$ relative to $r_\mathrm{c}$.
For $r \gg r_\mathrm{s}$ and $r \gg r_\mathrm{c}$, and $n=1$, the density falls of as $r^{-4}$. 

This profile has been successfully used to constrain projected halo mass profiles from observations of multiple images of gravitationally lensed background sources, see e.g.\@ \cite{bib:Kneib}, \cite{bib:Natarajan}, \cite{bib:Caminha}.
While \cite{bib:Kassiola}, \cite{bib:Natarajan}, and \cite{bib:Limousin} use Equation~\ref{eq:PIEMD_rho} to describe the dark matter halos of elliptical galaxies, \cite{bib:Kneib} and \cite{bib:Caminha}, among many others, also use it for galaxy-cluster scale mass density profiles.


\subsubsection{Jaffe profile}
\label{sec:Jaffe} 

The fifth commonly used dark matter halo profile, most often employed for galaxy-scale dark matter halos, is the Jaffe profile (JP), as introduced in \cite{bib:Jaffe}
\begin{equation}
\rho_\mathrm{J}(r) = \dfrac{\rho_\mathrm{s}}{\left( r/r_\mathrm{s} \right)^2 \left( 1 + r/r_\mathrm{s} \right)^{2}} \;.
\label{eq:Jaffe_rho}
\end{equation}
As for the other profiles, $\rho_\mathrm{s}$ and $r_\mathrm{s}$ denote the scaling density and scale radius, respectively.
For $r \ll  r_\mathrm{s}$, the density in the halo centre scales like $r^{-2}$, i.e.\@ even steeper than the MP, and it falls off like $r^{-4}$ for $r \gg r_\mathrm{s}$, similar to the PIEMD.
Yet, contrary to the latter, Equation~\ref{eq:Jaffe_rho} has a cuspy core, given by the first bracket in the denominator, and only one scale radius $r_\mathrm{s}$.

Projecting the JP on the two-dimensional sky, the de-Vaucouleurs profile, \cite{bib:Vaucouleurs}, is obtained. 
Indeed, Equation~\ref{eq:Jaffe_rho} is motivated as a corresponding three-dimensional mass density profile for the the projected, observed light distributions of elliptical galaxies and bulges of spiral galaxies that are heuristically fitted very well by de-Vaucouleurs profiles.

\section{Halo mass density profiles from a statistical viewpoint}
\label{sec:derivation}

Having introduced the physical mass density profiles, we now leave the mass-density approach until Section~\ref{sec:transfer} and pursue the statistical approach to describe a dark matter halo as an ensemble of particles.
The prerequisites for the calculations performed in Sections~\ref{sec:derivation} and \ref{sec:approximations} are stated in Section~\ref{sec:prerequisites}. 
Subsequently, we introduce the probability distribution to describe the location of an individual particle in a cosmic structure in Section~\ref{sec:Lomax}. 
The joint probability density distribution for an ensemble of identically distributed particles is set up in Section~\ref{sec:alpha}. 
We obtain the general equation to determine the power-law index for the one-particle probability distributions from the extrema of the joint distribution of the ensemble.

\subsection{Prerequisites}
\label{sec:prerequisites}

As a general starting point, we consider $n_\mathrm{p}$ independent, identical, and identically distributed particles of mass $m$, located at positions $\boldsymbol{r}_j$, $j=1,...,n_\mathrm{p}$ in a gravitationally bound structure.
We restrict ourselves to the case of a spherically symmetric structure (see Section~\ref{sec:profiles}).
Furthermore, we assume that the particles are only interacting with each other by weak-field gravity.
By the latter, we mean any gravitational law that can be effectively described as a scale-free interaction on a constant number of individual particles, which is also the basis of the hydrodynamical approach, the simulations, and the kinetic field theory of \cite{bib:Bartelmann_KFT} because observations of the large-scale matter power spectrum support this assumption, see e.g. \cite{bib:Planck}.
A good example fulfilling these requirements would be an ensemble of dark matter particles that interact with each other by Newtonian gravity to form a dark matter halo.

Here, the meaning of ``particle" depends on the scale that we want to describe the structure on. 
In simulations, for instance, ``particles" can be dark matter agglomerations of several thousand solar masses. 
In observations, the smallest ``particles" that can be described are given by the resolution at which the dark matter halo is constrained given all measurement imprecisions and uncertainties of the modelling.

The assumption of independent particles implies that they are uncorrelated. 
This is \com{similar to approaches mentioned in \cite{bib:Hjorth}.}
It is a reasonable starting point because any correlation between the particles requires additional interactions \com{and} evidence for additional dark matter particle interactions apart from gravitation is still to be observed.
\com{Including correlations} requires additional knowledge of the dynamical status of the structure such that correlations are traced back to a previous state of the structure -- which we do not take into account in this work. 
As we will show in the subsequent parts of the series, this assumption seems to be a good approximation for current simulations and observations.

Whether the particles in the ensemble are distinguishable or not, is of no concern.
We will see in the course of the calculations that our approach treats distinguishable and indistinguishable particles alike.
The equations to constrain the power-law index are invariant with respect to this property.

\subsection{The localisation probability of a particle within a structure}
\label{sec:Lomax}

As already stated in Section~\ref{sec:prerequisites}, following the principle of indifference, we assume that the identical particles are identically distributed. 
Given that the interaction which is supposed to generate the halo shape is scale-free, let
\begin{equation}
p(r) = N \left(1+\dfrac{r}{r_\sigma} \right)^{-(\alpha+1)}
\label{eq:p}
\end{equation}
be the probability of a single particle to be located at radial position $r$ in a three dimensional, spherically symmetric mass density distribution. 
To obtain a dimensionless radius, we scale $r$ by a scale radius $r_\sigma $ ($0 < r_\sigma < \infty$). 
From the mathematical point of view, the scaling to a dimensionless radius is required for the log-likelihood of the joint probability distribution of the entire ensemble.
$\alpha$ is the power-law index. 
We require $\alpha > -1$ to ensure that the probability density distribution monotonically decreases with increasing distance from the halo centre.

\com{We are free to choose $r_\sigma$ such that $r > r_\sigma$ for all particles.
This allows us to drop the 1 in Equation~\ref{eq:p} because, $r > r_\sigma$ avoids the singularity of a standard power-law at $r=0$. 
$p(r)$ thus remains finite and well-defined.}
We determine the normalisation constant $N$ from the requirement that the particle belonging to the structure is located in the volume that is delimited by the minimum radius, $r_\mathrm{min}$, and maximum radius of the structure, $r_\mathrm{max}$ ($r_\mathrm{max} > r_\mathrm{min} > r_\sigma$), i.e.\@
\begin{equation}
4 \pi \int \limits_{r_\mathrm{min}}^{r_\mathrm{max}} p(r) \, r^2 \mathrm{d}r = 4 \pi N \int \limits_{r_\mathrm{min}}^{r_\mathrm{max}} r^2 \left(\dfrac{r}{r_\sigma}\right)^{-(\alpha+1)} \mathrm{d}r \stackrel{!}{=} 1 \;.
\label{eq:N}
\end{equation}
Solving Equation~\ref{eq:N} as detailed in Appendix~\ref{app:N} we arrive at
\begin{equation}
N = \dfrac{1}{4 \pi r_\sigma^3} \dfrac{1}{g(\alpha, x_\mathrm{max}, x_\mathrm{min})}\;,
\label{eq:N_sol}
\end{equation}
in which, using the abbreviation $x_\mathrm{max} \equiv r_\mathrm{max}/r_\sigma$  \com{(and analogously for $x_\mathrm{min}$)},
\begin{equation}
g(\alpha, x_\mathrm{max}, x_\mathrm{min}) \equiv  \dfrac{x_\mathrm{max}^{2-\alpha}-x_\mathrm{min}^{2-\alpha}}{2-\alpha} \;.
\label{eq:f}
\end{equation}
Hence, the normalisation constant is a function of $\alpha$, $r_\sigma$, $r_\mathrm{min}$, and $r_\mathrm{max}$. 
Combining Equations~\ref{eq:p} and \ref{eq:N_sol}, we obtain the normalised probability distribution
\begin{equation}
p(r) =  \dfrac{1}{4 \pi r_\sigma^3} \dfrac{1}{g(\alpha, x_\mathrm{max}, x_\mathrm{min})} \left(\dfrac{r}{r_\sigma} \right)^{-(\alpha+1)}
\label{eq:Lomax}
\end{equation}
for a particle being localised at radius $r$ within the structure.

\subsection{The power-law index of the most likely ensemble configuration}
\label{sec:alpha}

Next, we determine the log-likelihood for the $n_\mathrm{p}$ independent and identically distributed particles as
\begin{equation}
\mathcal{L}(\alpha) = \ln \left( \prod \limits_{j=1}^{n_\mathrm{p}} p(r_j) \right) = \sum  \limits_{j=1}^{n_\mathrm{p}} \ln \left( N \left(\dfrac{r_j}{r_\sigma} \right)^{-(\alpha+1)}  \right) \;.
\label{eq:likelihood}
\end{equation}
The maximum-likelihood estimator for $\alpha$ is then determined by solving $\mathrm{d} \mathcal{L}/\mathrm{d} \alpha = 0$.
From Equation~\ref{eq:likelihood}, we obtain
\begin{equation}
\dfrac{\mathrm{d} \mathcal{L}(\alpha)}{\mathrm{d} \alpha} =  - \dfrac{\tfrac{\mathrm{d}g(\alpha, x_\mathrm{max}, x_\mathrm{min})}{\mathrm{d} \alpha}}{g(\alpha, x_\mathrm{max}, x_\mathrm{min})} - \dfrac{1}{n_\mathrm{p}} \sum \limits_{j=1}^{n_\mathrm{p}} \ln \left( \dfrac{r_j}{r_\sigma} \right) = 0 \;,
\label{eq:mle}
\end{equation}
as derived in detail in Appendix~\ref{app:mle}.
\com{Using Equation~\ref{eq:f}, we obtain for the first term in Equation~\ref{eq:mle}
\begin{align}
\dfrac{\tfrac{\mathrm{d}g(\alpha,x_\mathrm{max},x_\mathrm{min})}{\mathrm{d} \alpha} }{g(\alpha,x_\mathrm{max},x_\mathrm{min})} =& \dfrac{1}{2-\alpha} - \dfrac{\ln(x_\mathrm{max}) x_\mathrm{max}^{2-\alpha} - \ln(x_\mathrm{min}) x_\mathrm{min}^{2-\alpha}}{x_\mathrm{max}^{2-\alpha} - x_\mathrm{min}^{2-\alpha}} \;.
\end{align}
Assuming that $x_\mathrm{max} \gg x_\mathrm{min}$ for $\alpha \ne 2$, we arrive at
\begin{equation}
\dfrac{\tfrac{\mathrm{d}g(\alpha,x_\mathrm{max},x_\mathrm{min})}{\mathrm{d} \alpha} }{g(\alpha,x_\mathrm{max},x_\mathrm{min})} \approx \left\{ \begin{matrix} \tfrac{1}{2-\alpha} - \ln(x_\mathrm{max})  &  \text{for} \; \alpha < 2 \\[1ex]  \tfrac{1}{2-\alpha} - \ln(x_\mathrm{min}) & \text{for} \, \alpha > 2 \end{matrix} \right. \;.
\label{eq:dgg_approx}
\end{equation}}

As mentioned in Section~\ref{sec:prerequisites}, we find that Equation~\ref{eq:mle} is the same for distinguishable and indistinguishable particles because indistiguishable particles only require a global factor of $1/n_\mathrm{p}!$ multiplied to the product of probabilities in Equation~\ref{eq:likelihood}.
As this factor is a constant, it does not occur in the derivative of $\mathcal{L}(\alpha)$ in Equation~\ref{eq:mle}.

In fact, Equation~\ref{eq:mle} yields an extremum of the likelihood and not necessarily its maximum. 
It thus remains to be determined whether the solution found is a likelihood minimum, saddle point, or maximum.

We note that $\alpha$ enters Equation~\ref{eq:mle} over the normalisation $N$, i.e. the first term, which is related to the predefined extent of the structure under consideration and does not depend on the distribution of the $n_\mathrm{p}$ particles.
This distribution is contained in the second term.
Although we described the probability of one particle to be at a certain location by a power law, the actual distribution of an ensemble of simulated or observed particles may deviate from this assumption. 
In Section~\ref{sec:approximations}, we therefore consider different actual distributions for the second term and investigate different approximations to Equation~\ref{eq:mle} to determine $\alpha$ for the most common physically motivated limiting cases.
For each of them, we also calculate, which extremum of the likelihood they represent.

\com{Analogously, we can derive Equation~\ref{eq:likelihood} with respect to $r_\sigma$, $r_\mathrm{min}$ and $r_\mathrm{max}$ to determine extremum configurations for the scaling radius and the boundaries of the structure. 
The derivations can also be found in Appendix~\ref{app:mle}.
For $r_\sigma$, we obtain,}
\begin{equation}
\dfrac{\mathrm{d} \mathcal{L}(r_\sigma)}{\mathrm{d} r_\sigma} = \dfrac{n_\mathrm{p}}{N(r_\sigma)} \dfrac{\mathrm{d}N(r_\sigma)}{\mathrm{d} r_\sigma} - (\alpha + 1) \sum \limits_{j=1}^{n_\mathrm{p}} \left(- \dfrac{1}{r_\sigma} \right) = \dfrac{n_\mathrm{p}}{r_\sigma} \left( -\alpha - 1 + (\alpha+1) \right) = 0 \;.
\label{eq:dLdrsigma}
\end{equation}
\com{This result implies there is no preferred scale radius $r_\sigma$, which is expected by construction, assuming that $r_\sigma$ was introduced as an auxiliary parameter to obtain dimensionless variables and assuming $r_j > r_\sigma$ for all particles $j=1, ..., n_\mathrm{p}$.}

\com{The calculation for $r_\mathrm{max}$ yields}
\begin{equation}
\dfrac{\mathrm{d} \mathcal{L}(r_\mathrm{max})}{\mathrm{d} r_\mathrm{max}} = \dfrac{n_\mathrm{p}}{N(r_\mathrm{max})} \dfrac{\mathrm{d}N(r_\mathrm{max})}{\mathrm{d} r_\mathrm{max}} = - \dfrac{n_\mathrm{p}}{g(\alpha, x_\mathrm{max}, x_\mathrm{min})} \dfrac{\mathrm{d} g(\alpha, x_\mathrm{max}, x_\mathrm{min})}{\mathrm{d} r_\mathrm{max}} \;.
\label{eq:dLdrmax}
\end{equation}
\com{With}
\begin{equation}
\dfrac{1}{g(\alpha, x_\mathrm{max}, x_\mathrm{min})} \dfrac{\mathrm{d} g(\alpha, x_\mathrm{max}, x_\mathrm{min})}{\mathrm{d} r_\mathrm{max}} = \left(2-\alpha \right) \dfrac{r_\mathrm{max}^{1-\alpha}}{r_\mathrm{max}^{2-\alpha} - r_\mathrm{min}^{2-\alpha}}
\label{eq:dLdrmax_g}
\end{equation}
\com{we conclude that, for arbitrary halo boundaries, the derivative of the likelihood with respect to $r_\mathrm{max}$ is always zero for $\alpha = 2$. 
Since $r_\mathrm{max} > r_\sigma > 0$, the last term in Equation~\ref{eq:dLdrmax_g} only approaches zero for $r_\mathrm{max} \rightarrow \infty$.
This result implies that, given a finite $r_\mathrm{max}$, like $r_{200}$, the only extremum is at $\alpha=2$.
The calculation for $r_\mathrm{min}$ is analogous and discussed in Appendix~\ref{app:mle} because $r_\mathrm{min}$ can simply be identified with the radial position of the innermost particle of the ensemble.}

\com{We can summarise the results so far that, as expected from a scale-free ansatz for $p(r)$, no preferred values for the scaling radius $r_\sigma$ or the halo boundaries $r_\mathrm{min}$ and $r_\mathrm{max}$ can be derived.
Therefore, we continue our analysis with Equation~\ref{eq:mle} to find extrema for the power-law index $\alpha$.}

\section{Power-law indices for different approximations}
\label{sec:approximations}

Equation~\ref{eq:mle} depends on the extension of the structure, given by the maximum \com{and minimum} radius, $r_\mathrm{max}$ \com{and $r_\mathrm{min}$}, the scale radius $r_\sigma$, the number of particles belonging to the structure, $n_\mathrm{p}$, and their radial positions $r_j$, $j=1,...,n_\mathrm{p}$.  
Thus, different approximations to Equation~\ref{eq:mle} are obtained \com{depending on the relationship between the pre-defined halo extent, $r_\mathrm{max}$ \com{and $r_\mathrm{min}$}, the scaling radius $r_\sigma$, and the radial particle positions $r_j$ with respect to each other and the number of particles belonging to the structure}.
In a simulation, the latter amounts to changing the resolution of the structures. 

\com{Having chosen $r_\sigma < r_\mathrm{min} \le r_j$, we} can then distinguish four different limiting cases. 
These four cases, subsequently discussed in individual sections, are:
\begin{enumerate}
\item The first case, we call the ``core case'', \com{considers the central halo part from $r_\mathrm{min}$ to the boundary of the core, called $r_\mathrm{core}$.} 
In this case, \com{we assume $r_\mathrm{min} < r_j < r_\mathrm{core} \ll r_\mathrm{max}$.
We furthemore introduce $n_\mathrm{c}$ as the number of particles in the core and assume that it is small compared to $n_\mathrm{p}$, the total amount of particles in the entire halo,} so that the individual particles move freely in the range between $r_\mathrm{min}$ and $r_\mathrm{core}$. 
\item The second case is the homogeneous fluid case, in which the particles are homogeneously distributed in the volume between $r_\mathrm{core}$ and $r_\mathrm{max}$.
This case requires large particle numbers up to $r_\mathrm{max}$, so that we consider $n_\mathrm{p} \rightarrow \infty$ here.

\item In the third case, we call the ``thinning outskirts case'', we \com{let} $r_\mathrm{max}$ \com{go} to large extensions, $r_\mathrm{max} \rightarrow \infty$, while the particles assigned to the structure are assumed to remain at a much smaller radius, $r_j \ll \infty$, $j=1,...,n_\mathrm{p}$.  
This case ignores particles in the simulated or observed particle ensemble that are gravitationally bound to a structure but are located far from its centre.
\item The fourth case is called the ``overflowing outskirts case'', in which we assume $r_\mathrm{max}$ is chosen too small, such that the particle distribution assigned to that structure spills over the predefined maximum extension. 
\end{enumerate}

\subsection{Core case}
\label{sec:core}

At first, we consider the \com{central part of the halo from $r_\mathrm{min}$ up to a radius $r_\mathrm{core}$ that we call the boundary of the halo core. 
Inspired by simulations, we take $r_\mathrm{min}$ to be the inner-most radius that is available, $r_\mathrm{core}$ to be the radius at which the simulation has converged and $r_\mathrm{max}$ to be $r_{200}$. 
For the simulations of \cite{bib:NFW3}, $r_\mathrm{min} = 1.5 \cdot 10^{-4} \, r_{200}$ and $r_\mathrm{core} = 0.6 \cdot 10^{-3} \, r_{200}$, such that $r_\mathrm{min}/r_\mathrm{core} = 2.7 \cdot 10^{-2}$.
We thus assume that $r_\mathrm{min} \ll r_\mathrm{core}$, so that we can use Equation~\ref{eq:mle} and insert Equation~\ref{eq:dgg_approx} with $r_\mathrm{core}$ now taking the role of $r_\mathrm{max}$ as the outer boundary of the central halo part to obtain\footnote{Knowing the result, we directly insert the appropriate case with $r_\mathrm{max}$ from Equation~\ref{eq:dgg_approx}.}
\begin{equation}
 - \left( \dfrac{1}{2-\alpha} - \ln(x_\mathrm{core}) \right) - \dfrac{1}{n_\mathrm{c}} \sum \limits_{j=1}^{n_\mathrm{c}} \ln \left( x_j \right) = 0 \;.
\end{equation}
which, using that $x_j/x_\mathrm{core} = r_j/r_\mathrm{core}$, can be simplified as 
\begin{equation}
\dfrac{1}{2-\alpha} = - \dfrac{1}{n_\mathrm{c}} \sum \limits_{j=1}^{n_\mathrm{c}} \ln \left( \dfrac{r_j}{r_\mathrm{core}} \right) = - \dfrac{1}{n_\mathrm{c}} \sum \limits_{j=1}^{n_\mathrm{c}} \ln \left(1 + \dfrac{r_j}{r_\mathrm{core}} -1 \right) \approx  1 - \dfrac{1}{n_\mathrm{c}} \sum \limits_{j=1}^{n_\mathrm{c}} \dfrac{r_j}{r_\mathrm{core}} \;.
 \label{eq:core}
\end{equation}
The extremum values of $\alpha$ are thus independent of $r_\sigma$ and only subject to the number of particles in the core, $n_\mathrm{c}$, and their distribution.
The accuracy of the approximation of the logarithm on the right-hand side increases, the closer $r_j$ is to $r_\mathrm{core}$. 
Since we assume that $0 < r_j < r_\mathrm{core}$, convergence of the full Taylor expansion is fulfilled.} 

\com{For a homogeneous distribution of particle radii between $r_\mathrm{min}$ and $r_\mathrm{core}$, we have 
\begin{equation}
\dfrac{1}{n_\mathrm{c}} \sum \limits_{j=1}^{n_\mathrm{c}} \dfrac{r_j}{r_\mathrm{core}} = \dfrac{1}{2} \left( 1 + \dfrac{r_\mathrm{min}}{r_\mathrm{core}} \right) \approx \dfrac{1}{2} \;,
\end{equation}
such that
\begin{equation}
\alpha = 2 - \left( 1 - \dfrac{1}{n_\mathrm{c}} \sum \limits_{j=1}^{n_\mathrm{c}} \dfrac{r_j}{r_\mathrm{core}} \right)^{-1}  = 0 \;.
\label{eq:alpha_core}
\end{equation}}

Assuming that it is not the radius which is uniformly distributed but the three-dimensional particle position in the volume of the halo, we obtain 
\begin{equation}
\dfrac{1}{n_\mathrm{c}} \sum \limits_{j=1}^{n_\mathrm{c}} \dfrac{r_j}{r_\mathrm{core}} = \dfrac{1}{r_\mathrm{core}} \dfrac{\tfrac34 r_\mathrm{core} - \left(  \tfrac{r_\mathrm{min}}{r_\mathrm{core}} \right)^3 \tfrac34 r_\mathrm{min}}{1 - \left(  \tfrac{r_\mathrm{min}}{r_\mathrm{core}} \right)^3}  \approx \dfrac{3}{4} \;,
\end{equation}
\com{such that $\alpha = -2$, which is already smaller than the smallest possible $\alpha=-1$.}

Hence, the specific particle distribution directly determines $\alpha$.
\com{For the uniform distribution of particle radii according to Equation~\ref{eq:alpha_core}, we find $p(r) \propto r^{-1}$.
Shifting the mean radius of the particle distribution towards $r_\mathrm{core}$ leads to shallower profiles, until we reach $p(r) \propto r^0$. 
In simulations, $p(r)$ is constant per bin, such that $p(r) \propto r^0$ occurs naturally as soon as $r_\mathrm{core} = r_\mathrm{min}$, when the core only covers the first bin filled with particles.
Similarly, probability densities steeper than $p(r) \propto r^{-1}$, i.e. $\alpha >  0$ are obtained when the mean radius of the particle distribution is shifted away from $r_\mathrm{core}$. Any value of $\alpha < 2$ given by Equation~\ref{eq:alpha_core} leads to a likelihood \com{maximum}.}

\subsection{Homogeneous fluid case}
\label{sec:homogeneous}

\com{Next, we describe intermediate part of the halo with the particle distribution from $r_\mathrm{core}$ up to the halo boundary $r_\mathrm{max}$.
Since $r_\mathrm{min}$ and $r_\mathrm{core}$ are both much smaller than $r_\mathrm{max}$, for the sake of simplicity, we consider the entire halo from $r_\mathrm{min}$ to $r_\mathrm{max}$.}

\com{As in Section~\ref{sec:core}, we insert Equation~\ref{eq:dgg_approx}} into Equation~\ref{eq:mle}
\begin{equation}
 - \dfrac{1}{2-\alpha} - \dfrac{1}{n_\mathrm{p}} \sum \limits_{j=1}^{n_\mathrm{p}} \ln \left( \dfrac{r_j}{r_\mathrm{max}} \right) = 0
\end{equation}
\com{and solve for $\alpha$ to obtain
\begin{equation}
\alpha = 2 + \dfrac{n_\mathrm{p}}{\sum \limits_{j=1}^{n_\mathrm{p}} \ln \left( \dfrac{r_j}{r_\mathrm{max}} \right)}  \;.
\label{eq:alpha_fluid}
\end{equation}
With $0 < r_\mathrm{min} \le r_j \le r_\mathrm{max} < \infty$, all terms in the denominator are negative yielding $\alpha < 2$.}

\com{Assuming that $n_\mathrm{p} \gg 1$, the particle distribution becomes a homogeneous fluid and the particles} are located at positions
\begin{equation}
r_j = \dfrac{j}{n_\mathrm{p}} r_\mathrm{max} \quad \Rightarrow  \quad \dfrac{r_j}{r_\mathrm{max}} = \dfrac{j}{n_\mathrm{p}} \;, \; j=1,...,n_\mathrm{p} \;.
\label{eq:hom_n}
\end{equation}
This implies that the power-law index of the \com{extremum} particle distribution becomes independent of $r_\mathrm{max}$ and only depends on $n_\mathrm{p}$.
Assuming that we narrow the radius range of a power law and cut off the strongly decreasing tail, we can understand the homogeneous fluid distribution as a detail of a power law distribution, so that the actual particle distribution can still be a power law. 

We approximate the sum in the denominator by Stirling's formula as detailed in Appendix~\ref{app:Stirling} to arrive at
\begin{equation}
\lim \limits_{n_\mathrm{p} \rightarrow \infty} \alpha(n_\mathrm{p}) = \lim \limits_{n_\mathrm{p} \rightarrow \infty} 2 - \dfrac{n_\mathrm{p}}{n_\mathrm{p} - \mathcal{O}(\ln(n_\mathrm{p}))} = 1 \;,
\label{eq:fluid_limit}
\end{equation}
which belongs to a likelihood maximum. 
Inserting this power-law index into Equation~\ref{eq:Lomax}, we arrive at a power-law distribution for the individual particles that scales with $r^{-2}$.

\subsection{Thinning outskirts case}
\label{sec:thinning_outskirts}

For the ``thinning outskirts case'', we reuse Equation~\ref{eq:alpha_fluid}. 
Instead of assuming a homogeneous distribution for the sum term in the denominator, we assume that
\begin{equation}
r_j  \ll r_\mathrm{max} \; \forall j=1,...,n_\mathrm{p} \quad \Rightarrow  \quad \sum  \limits_{j=1}^{n_\mathrm{p}} \ln \left( \dfrac{r_j}{r_\mathrm{max}} \right) = \sum  \limits_{j=1}^{n_\mathrm{p}} \ln \left( \dfrac{x_j}{x_\mathrm{max}} \right) \approx - n_\mathrm{p} \ln(x_\mathrm{max} )\;,
\label{eq:sum_tmax}
\end{equation}
\com{in which we have to reintroduce $x_\mathrm{max} = r_\mathrm{max}/r_\sigma$ again to make the argument of the logarithm dimensionless.}
This approximation, $\ln(x_j) \approx 0$ for all $j=1,...,n_\mathrm{p}$, implies that we do not take into account particles in the tail of the power-law distribution. 
These particles are gravitationally bound to the structure but may not be assigned to this structure because they are far away from its centre. In simulations or observations, they may also be included into neighbouring structures to which they are closer.

Inserting Equation~\ref{eq:sum_tmax} into Equation~\ref{eq:alpha_fluid}, we obtain
\begin{equation}
\alpha(x_\mathrm{max}) = 2 - \dfrac{1}{\ln(x_\mathrm{max})} \;.
\end{equation}
Letting $r_\mathrm{max}$ approach the range of the gravitational interaction length, we obtain
\begin{equation}
\lim \limits_{x_\mathrm{max} \rightarrow \infty} \alpha(x_\mathrm{max}) = 2 \;,
\label{eq:outskirts_limit}
\end{equation}
which results in a likelihood minimum and a one-particle power-law distribution that scales with $r^{-3}$.

\subsection{Overflowing outskirts case}
\label{sec:overflowing_outskirts}

Lastly, we treat the ``overflowing outskirts case'' by reusing Equation~\ref{eq:alpha_fluid} again and assuming that the volume of the structure is set too small, such that $r_\mathrm{max} \le r_j$ for all $j = 1,...,n_\mathrm{p}$, i.e. the particle distribution spills \com{completely} over the defined extension of the structure. 
Furthermore, we assume that the $r_j$ and $r_\mathrm{max}$ are of the same order of magnitude, such that 
\begin{equation}
\langle r \rangle = r_\mathrm{max}  + \langle \delta r \rangle \;, \quad \text{with}  \quad  \langle \delta r \rangle \le r_\mathrm{max}
\label{eq:means}
\end{equation}
holds for the arithmetic mean value $\langle r\rangle$ of the radial positions $r_j$ and the arithmetic mean $\langle \delta r \rangle$ of their deviations $\delta r_j$ from $r_\mathrm{max}$.

Given these assumptions, further motivated below, we can now approximate the sum in the denominator of Equation~\ref{eq:alpha_fluid} by using Equation~\ref{eq:means}
\begin{equation}
\sum  \limits_{j=1}^{n_\mathrm{p}} \ln \left( \dfrac{r_j}{r_\mathrm{max}} \right) = \sum  \limits_{j=1}^{n_\mathrm{p}} \ln \left( 1 + \dfrac{\delta r_j}{r_\mathrm{max}} \right) \approx \sum  \limits_{j=1}^{n_\mathrm{p}} \dfrac{\delta r_j}{r_\mathrm{max}} \equiv n_\mathrm{p}\, k \;.
\label{eq:sum_tmax2}
\end{equation}
In the last step, we defined $0 < k \le 1$.
Thus, we obtain
\begin{equation}
\alpha(k) = 2 + \dfrac{1}{k} \;.
\label{eq:alpha_outskirts2}
\end{equation}
Letting $k$ approach the maximum value, we obtain
\begin{equation}
\lim \limits_{k \rightarrow 1} \alpha(k) = 3 \;,
\label{eq:outskirts_limit2}
\end{equation}
which results in a likelihood maximum and a one-particle power-law distribution that scales with $r^{-4}$.

If we do not impose the assumptions stated in Equation~\ref{eq:means}, but instead assume that $r_j \gg r_\mathrm{max}$ for all $j=1,...,n_\mathrm{p}$, the sum in Equation~\ref{eq:sum_tmax2} can still be set equal to $n_\mathrm{p} \, k$, now for $k \gg  1$. 
Taking the limit of $k \rightarrow \infty$ implies that the arithmetic mean of the actual particle distribution goes to infinity for a finite $r_\mathrm{max}$.
We then obtain the same limit as in Equation~\ref{eq:outskirts_limit} again.
This case may occur, if it is very hard to separate the structure of interest from its environment and its volume is assumed too small compared to the actual extend of the structure.
Comparing this case with the thinning outskirts case, we thus arrive at consistent results, as we do not account for particles far away from the centre which should be assigned to the structure in both cases. 
If we can set $r_\mathrm{max}$ at least at the right order of magnitude with the second constraint in Equation~\ref{eq:means}, we can read off Equation~\ref{eq:alpha_outskirts2} that steeper one-particle power-law distributions than $r^{-4}$ are obtained, if $0 < \langle \delta r \rangle < r_\mathrm{max}$.
Hence, analogously to the core case (Section~\ref{sec:core}), the specific distribution of particles with respect to $r_\mathrm{max}$ fixes $\alpha$ in the boundary regions of a structure.

\com{Since $\alpha = 3 > 2$, the approximation of Equation~\ref{eq:dgg_approx} using $\ln(r_\mathrm{max})$ seems inappropriate to use here. Yet, requiring that the particle distribution is located at radii all larger than $r_\mathrm{max}$, the condition that $r_\mathrm{max} > r_\mathrm{min}$ used in this approximation is also inverted. We can thus continue to employ this approximation, as it remains valid for this case.}

\section{From particle ensembles to mass density profiles}
\label{sec:transfer}

In statistical mechanics, the transfer from the positions and momenta of an ensemble of $n_\mathrm{p}$ particles to continuous spatial number densities and densities in momentum space is induced by defining a probability distribution function in phase-space.
This function assigns each element of the phase-space volume a probability that the configuration of the positions and momenta of the particle ensemble currently under consideration lies in this phase-space volume-element, see e.g.\@ \cite{bib:Bartelmann} for an introduction.
The size of the infinitesimal phase-space volume-element thus sets a scale below which individual particle properties are averaged over.
For a collisionless system of particles, the individual one-particle probability density functions in phase-space are uncorrelated, which implies that the probability distribution function of the ensemble is separable and amounts to a product of one-particle probability density functions, see \cite{bib:Binney}.
Thus the prerequisites of collisionless systems are consistent with our prerequisites in Section~\ref{sec:prerequisites}, especially the assumption that the particles are independent, and hence allow us to link both approaches. 

In the framework of phase-space probabilities, the number density of particles at a position $r$, $n(r)$, is given as the probability density function in phase-space for one particle integrated over all its potential momenta and normalised to obtain all $n_\mathrm{p}$ particles when integrating over the entire volume. Consequently, the number density of particles can be determined from the one-particle probability density function of Equation~\ref{eq:p} as
\begin{equation}
n(r) = n_\mathrm{p} \, p(r) \;,
\label{eq:n}
\end{equation}
in which $n_\mathrm{p}$ enters due to the normalisation condition mentioned above.

Using Equation~\ref{eq:n}, it is straighforward to determine the mass density of a structure consisting of identical particles with mass $m$ that are distributed according to the number density $n(r)$ as
\begin{equation}
\rho(r) = m n(r) = m n_\mathrm{p}\, N(\alpha,r_\sigma,r_\mathrm{max},r_\mathrm{min}) \left( \dfrac{r}{r_\sigma} \right)^ {-(\alpha+1)} \equiv \rho_\sigma  \left( \dfrac{r}{r_\sigma} \right)^ {-(\alpha+1)} \;.
\label{eq:rho_n}
\end{equation}
In the last step, we introduced a scaling density $\rho_\sigma$ in analogy to the scaling density $\rho_\mathrm{s}$ for the mass densitiy profiles of Section~\ref{sec:profiles}.
Integrating $\rho(r)$, we obtain the mass of the structure up to radius $r$
\begin{equation}
M(r) = m n_\mathrm{p}\, 4 \pi \int \limits_{r_\mathrm{min}}^{r} \mathrm{d} \hat{r} \,  \hat{r}^2 p(\hat{r}) \;.
\label{eq:halo_mass}
\end{equation}

\begin{figure*}[t]
\begin{center}
  \includegraphics[width=0.66\textwidth]{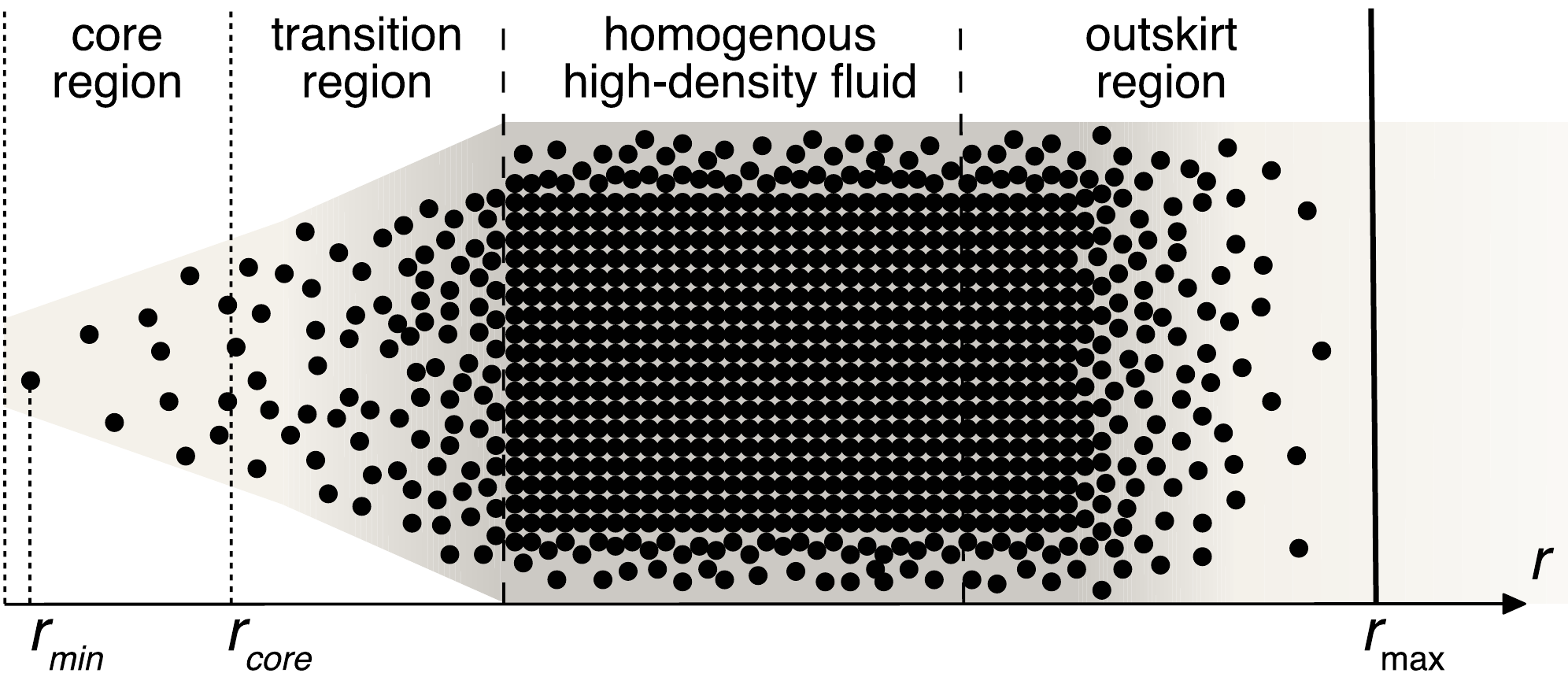}
\caption{Schematic composition of the most likely dark matter halo shape based on the probabilistic derivations and approximations detailed in Section~\ref{sec:derivation} (here shown with ``thinning outskirts'').}
\label{fig:halo}       
\end{center}
\end{figure*}

As a result from this transfer, we find that the mass density profile of a structure $\rho(r)$ consisting of an ensemble of collisionless particles is directly proportional to the one-particle probability distribution $p(r)$ as introduced in this section.
Consequently, we can assemble a dark matter halo in its most likely shape as depicted in Figure~\ref{fig:halo} out of an ensemble of dark matter particles fulfilling the prerequisites of Section~\ref{sec:prerequisites}. 

\paragraph{The core:} The innermost part of the halo is given by radial positions that are smaller than \com{some chosen core radius, called $r_\mathrm{core}$}. 
There, dark matter forms an ideal gas \com{of collisionless freely moving particles}.
\com{The slope of the corresponding} mass density profile is determined by the specific distribution of the particles. 
The arithmetic mean radius around which the particle distribution is concentrated can be employed as its characteristic length.
If the radii of the particles are uniformly distributed, a $\rho(r) \propto 1/r$-behaviour is observed (see Section~\ref{sec:core}).
Shallower mass density profiles are expected for a mean radius closer to $r_\mathrm{core}$, steeper profiles for a mean radius closer to the \com{halo centre}.  
\\[-0.65cm]
\paragraph{The homogeneous fluid region:} With increasing radial distance from the halo centre, the dark matter particles become more numerous and dense to form a homogeneous fluid for the limiting case of infinitely many particles (see Section~\ref{sec:homogeneous}).
The same limiting $\rho(r) \propto 1/r^2$-behaviour as for this region is also derived for the mass density profile of a singular isothermal sphere using the equation of hydrostatic equilibrium for an ideal gas density, see e.g.\@ \cite{bib:Binney} for details.
The former arises as the most likely configuration that any self-gravitating system of independent particles strives towards.
The latter is the maximum-entropy configuration of the gas in phase space, \cite{bib:Bartelmann}. 
Thus, the ideal isothermal gas with its smooth mass density profile has the same properties as our fluid approximation of the collisionless dark matter particle ensemble.
In this sense, our mathematical derivation of $\alpha$ from the most likely spatial configuration of an ensemble of particles in Section~\ref{sec:homogeneous} based on probability theory is consistent with the physical one based on statistical mechanics and thermodynamics.
\\[-0.65cm]
\paragraph{The boundary region (``outskirts''):} 
How the structure is separated from others or from a background, strongly depends on the definition of the structure and its specific environment. 
As no permanent thermodynamical equilibrium states exist due to the infinite range of the gravitational interaction, the definition of a maximum radius to separate structures is arbitrary compared to standard definitions based on equilibrium states. 
Consequently, depending on the location at which we set $r_\mathrm{max}$ with respect to the ensemble mean of the particle distribution that forms the halo of interest, we obtain different slopes for the decreasing power-laws of $\rho(r)$ (see Sections~\ref{sec:thinning_outskirts} and \ref{sec:overflowing_outskirts}). 
\com{As Equation~\ref{eq:dLdrmax} reveals, $\rho(r) \propto r^{-3}$ is preferred to obtain an extremum of the likelihood for any chosen $r_\mathrm{max} < \infty$.}
In any of those cases, the derivations show that the respective power-law slope is set by discarding weakly-bound particles far from the halo centre.

\subsection{Explanation for the most commonly used mass density profiles}
\label{sec:matching}

\begin{table}
\caption{Integration of all models of Section~\ref{sec:profiles} into our approach using the approximations as detailed in Section~\ref{sec:approximations} to Equation~\ref{eq:mle}. Section~\ref{sec:matching} contains further details.}
\label{tab:summary_models}       
\begin{tabular}{llll}
\hline\noalign{\smallskip}
mass density  & core & outskirts & comments \\
profile & $\alpha(k) = 2 -1/k$ &  & \\
\noalign{\smallskip}\hline\noalign{\smallskip}
NFW & $k=1/2$ & thinning & uniform radial distribution of particles in the core\\
MP & $k \boldsymbol{<} 1/2$ & thinning & steeper radial distribution of particles in the core\\
EP ($\beta =0$) & $k=1$ & no outskirts & this case is an isothermal sphere \\
PIEMD & no power-law & overflowing & inner part is not a power-law distribution \\
JP & $ k = 1$ &  overflowing & core is an isothermal fluid \\
\noalign{\smallskip}\hline
\end{tabular}
\end{table}

As a summarising result of this section, we disentangled a dark matter halo into a core, an intermediate, and an outskirts part, based on a mass density profile that is proportional to the power law in Equation~\ref{eq:p}.
Comparing the extremal power-law indices $\alpha$ for these halo parts, we find that they vary with the distance from the halo centre and decisively depend on the predefined halo extensions and given morphological symmetry.  
In addition, $\alpha$ depends on the underlying distribution of dark matter particles.

Hence, we can now comprehend why the most commonly used dark matter halo mass density profiles of Section~\ref{sec:profiles} are good fits to most of the simulated dark matter halos and those inferred from observations:
All heuristic models are able to produce the varying $\alpha$ over the extension of the halo. 
Starting at the core, the profile slope steepens with increasing radius, such that $\alpha$ monotonically increases with increasing radius. 
For the NFW profile, the MP, and the JP, the mass density profile consists of a product of two power laws which can be directly related to the approximating cases of Equation~\ref{eq:mle} as summarised in Table~\ref{tab:summary_models}.
These profiles thus seem to be constructed based on the physically motivated prerequisites stated in Section~\ref{sec:prerequisites}, in particular on the assumption that the scale-free gravitational interaction can be described by a power-law probability density.
The varying $\alpha$s between these models account for the fact that different approximations to Equation~\ref{eq:mle} lead to slightly varying extremal power-law indices, as detailed in Section~\ref{sec:approximations}. 
Simulation- and observation-based examples for these approximations will be shown in the next parts of the series.

Potential reasons for the debates about the profile slope in the core and in the outskirts part can now be explained by the result that the extremal power-law index for \com{these parts are highly dependent on the specific particle distribution. 
In addition, the extremum of the thinning outskirts case is a likelihood minimum.} 
Contrary to that, all density profiles agree that halos show an isothermal part between the core and the outskirts, which belongs to the most likely spatial configuration of a many-particle ensemble and is therefore ubiquitous in simulated and observed cosmic structures.
The overflowing outskirts is also a likelihood maximum. 
Its decrease of the mass density profile is often found for galaxy-scale masses. 
On that scale, the extension of the structure is characterised by the half-light radius, which seems to be a good estimate for the order of magnitude on which the mass is distributed as well and which motivated the approximations stated in Equation~\ref{eq:means}.

For the EP and the PIEMD, the integration into our approach is different, as they are not based on the power law of Equation~\ref{eq:p}.
In order to find their extremal $\alpha$, the calculations of Sections~\ref{sec:derivation} and \ref{sec:approximations} need to be repeated using their specific probability density distributions.
Instead of doing this, we show that the two density profiles can approximate the power law of Equation~\ref{eq:p}, such that the EP and the PIEMD have a similar behaviour as the other profiles.
Consequently, this corroborates their good fits to simulations and observations of dark matter dominated structures.
Further examples for successful fits to simulations and observational data will follow alongside the ones mentioned above in the next parts of this series.

EPs for $\beta \ll 1$ approximate an isothermal sphere because the limit of $\beta \rightarrow 0$ for Equation~\ref{eq:Einasto_rho} yields
\begin{equation}
\rho_\mathrm{E}(r) = \dfrac{\rho_\mathrm{s} }{(r/r_\mathrm{s})^2} \;.
\end{equation}
For $\beta \ne 0$ and $r \ll r_\mathrm{s}$, the exponential function can be Taylor-expanded to yield 
\begin{equation}
\rho_\mathrm{E} \propto \left(1 - \tfrac{2}{\beta} \left(\tfrac{r}{r_\mathrm{s}}\right)^\beta \right) \;,
\end{equation}
such that the core of an EP is similar to a power law.

For the PIEMD, the profile is a product of two different power laws constructed to yield the same behaviour as the power-law profiles mentioned above for large $r$. 
For $r \rightarrow 0$, it avoids any singularity and given $r \ll r_\mathrm{s} \le r_\mathrm{c}$, we obtain to leading order
\begin{equation}
\rho_\mathrm{PIEMD}(r) \propto \left(1 - \left(\tfrac{r}{r_\mathrm{s}}\right)^2 \right) \;,
\end{equation}
which is similar to the next-to-leading-order Taylor expansion of Equation~\ref{eq:p} without the linear term in $r$.

\section{Conclusion}
\label{sec:conclusion}

The main aim behind this new approach is to complement current models of structure growth that intertwine the description of a cosmic structure with its dynamics and thus require to track initial seed structures through cosmic time in order to explain the non-linear mass agglomerations we observe today.  
In the first part of this series on structure evolution over cosmic time, we focussed on separating the description of a cosmic structure from its dynamics and reducing the former to the necessary amount of prerequisites and characterising equations. 
Being able to find a description of any structure at an arbitrary point in cosmic time that can be used as a starting point for any theory of structure growth is a first step to making our standard approaches more efficient and to envisioning tests of alternatives that have been computationally intractable so far.

Our goal in this first part was to derive a general description of dark matter halo shapes from more fundamental principles than using heuristic models like the ones introduced in Section~\ref{sec:profiles}.
Our newly gained understanding of the morphology of dark matter halos also allows us to support the currently employed gravitational lensing models to infer the Hubble-Lemaître constant (see e.g.~\cite{bib:Millon} and references therein for recent investigations of potentially arising biases in the modelling) or to infer galaxy cluster masses (see e.g.~\cite{bib:Meneghetti} and references therein for recent investigations on the quality of mass density reconstructions and potentially arising biases in the modelling). 

To describe the morphology of a dark matter halo, we employed the physical prerequisites of a set of collisionless, identical, only gravitationally interacting particles that are confined to a spherically limited volume called the dark matter halo. 
Translating these requirements into mathematical terms yields an ensemble of independent, identical particles that are identically distributed within a finite sphere according to a power law probability density distribution. 
We set up the joint probability density distribution of the entire ensemble and determined the power-law indices that belong to the most and least likely spatial configurations of the ensemble, given certain physically motivated limits on the confining volume and the number of particles in the ensemble, as detailed in Section~\ref{sec:approximations}. 

\com{In agreement with previous works, e.g.~\cite{bib:NFW1}, we find that the scaling radius $r_\sigma$ introduced to obtain a dimensionless radial variable drops out of the equation that constrains the power-law index. 
Due to the scale-free approach, there is no preferred $r_\sigma$ (see Equation~\ref{eq:dLdrsigma}). 
Nevertheless, introducing $r_{-2}$, the radius at which the slope of the mass density profile is $-2$, as scaling radius, the concentration parameter $c=r_{-2}/r_{200}$ yields valuable information about the scale of isothermality with respect to the scale of halo relaxation.
Since isothermality is reached with an infinite number of dark matter particles in our approach, while all other parts contain a finite amount of freely moving particles, we will further investigate whether a $r_\sigma$ exists as a phase-transition scale between a dark matter fluid and freely moving particles. A succeeding investigation can characterise the mean free path of (potentially colliding) dark matter particles and allows to interpret the halo concentration parameter as the Knudsen number for dark matter halos.} 

\com{Beyond existing approaches, we arrive at a derivation of the power-law index of the mass density profiles without the need to employ any equilibrium considerations and the velocity distribution of the particle distribution. 
Avoiding the combination of dynamics and structure description, our approach clearly shows the connection between a joint, spatial configuration of dark matter particles on the microscopic scale and their respective mass density profile as their averaged macroscopic representation.}
As a result, we find that the mass density profile belonging to a set of freely moving particles confined to a \com{small} volume \com{in the inner-most halo part} shows the same behaviour as assumed in the core region of the heuristic power-law mass density profiles introduced in Section~\ref{sec:profiles}.
The fact that the extremum power-law index \com{strongly depends on the specific distribution of the particles in the core with respect to the chosen maximum core radius}, could resolve the cusp-core-debate (see \cite{bib:Bullock} for an introduction) without the need to introduce any effects caused by luminous matter, as further detailed in Section~\ref{sec:matching}. 
Analogously, we succeeded in retrieving the isothermal intermediate part between core and outskirts in our framework. 

When determining the power-law indices for the outskirts on galaxy- and galaxy-cluster scale (see also Section~\ref{sec:matching} for further details), we found that the power-law index is set by the relation between the predefined halo boundary and the maximum radius of the particle ensemble under consideration. 
This result enables us to understand why the galaxy- and galaxy-cluster scale halos show a different decrease in the mass density profile far from the centre.
It emerges due to the two different definitions of halo boundaries.
On galaxy scale, we observe the half-light radius of the luminous part of the galaxy and assume that the respective dark matter halo is of approximately the same order of magnitude in its extensions. 
For galaxy-clusters, which we consider the largest cosmic structures that form the nodes in the cosmic web, we assume that, in principle, all particles irrespective of their distance to the cluster centre are gravitationally bound to the halo.
At the same time, we divide the total ensemble of particles in a simulation or observation on large scales into several neighbouring clusters, such that the mass density profile for an individual cluster arises as the trade-off between the long-range effect of gravitational attraction and the assignment of particles to their nearest local minimum of the gravitational potential.

On the whole, our approach shows that an individual dark matter halo can be considered an \emph{emergent} structure, i.e. an assignment of individual particles to an ensemble of particles close to a local minimum in the gravitational potential according to predefined boundary and symmetry constraints.
Hence, our picture of a halo is a transient state of a set of particles agglomerating around a salient point in space.
Over cosmic time, different particles can be assigned to the halo, such that its shape can be maintained with a different set of particles or its morphology can change as a consequence of the dynamics, for instance allowing for merger events.

Since we aim at describing individual halos as required by the observational applications like gravitational lensing or kinematical characterisations, our halo model is currently not linked to the power spectrum of density fluctuations, which is the basis of the approach outlined in \cite{bib:Bartelmann_KFT}.

Furthermore, it is different from the current standard model based on fluid dynamics, as already mentioned in the introduction. 
In this picture, halos are \emph{entities} of a specified set of particles that is tracked in their collective motion over cosmic time, such that the particle dynamics introduces changes in the halo morphology. 

Given its success at explaining the dark matter halo shapes in a very clear and simple way, the subsequent parts of this series will deal with a thorough consistency check of the proposed approach by dark-matter-only simulations and by observations of low surface brightness galaxies, which are assumed to mainly consist of dark matter. 
Beyond that, we will also apply our approach to corroborate the common mass density models for gravitational lenses, see e.g.~\cite{bib:Wagner3} for an overview, and investigate whether it can shed light onto the bulge-halo-conspiracy, \cite{bib:Auger}. 
Thereby, we intend to integrate our approach into the landscape of existing methods to improve our understanding of structure growth and enhance our tool set for doing so.

%

\begin{acknowledgements}
I thank George F.~R.~Ellis and Carlo Rovelli for their inspiring works leading to this approach. In addition, I thank Xingzhong Er, Robert Grand, Jiaxin Han, Bettina Heinlein, \com{Jens Hjoth}, Sebastian Kapfer, Angela Lahee, Christophe Pichon, Andrew Robertson, Johannes Schwinn, Volker Springel, R\"{u}diger Vaas, Gerd Wagner, \com{Liliya Williams, and the anonymous referee} for helpful comments, as well as the participants of the First Shanghai Assembly on Cosmology and Galaxy Formation 2019 for many helpful discussions and encouragement to further pursue this idea. \com{I also like to thank Andrea Macciò for fruitful discussions and support and the New York University Abu Dhabi for their hospitality.} I gratefully acknowledge the support by the Deutsche Forschungsgemeinschaft (DFG) WA3547/1-3.
\end{acknowledgements}

%
%

\bibliographystyle{spmpsci}      
\bibliography{aa}   


\appendix

\section{Derivation of $g(\alpha,r_\sigma,r_\mathrm{max},r_\mathrm{min})$}
\label{app:N}

To determine the integral
\begin{equation}
I(\alpha, r_\sigma, r_\mathrm{max},r_\mathrm{min}) = \int \limits_{r_\mathrm{min}}^{r_\mathrm{max}} r^2 \left(\dfrac{r}{r_\sigma}\right)^{-\alpha-1} \mathrm{d}r \;,
\end{equation}
we substitute
\begin{equation}
x = \tfrac{r}{r_\sigma} \;, \quad \Leftrightarrow \quad r = r_\sigma x \;, \quad \Rightarrow \quad \mathrm{d}r = r_\sigma \mathrm{d}x \;.
\end{equation}
and thus obtain
\begin{align}
I(\alpha, r_\sigma, r_\mathrm{max},r_\mathrm{min}) &= \int \limits_{r_\mathrm{min}}^{r_\mathrm{max}} r^2 \left(\dfrac{r}{r_\sigma}\right)^{-\alpha-1} \mathrm{d}r  \\
&=  r_\sigma^3 \int \limits_{x_\mathrm{min}}^{x_\mathrm{max}} \left( x^{1-\alpha}\right) \mathrm{d} x \\
&=  r_\sigma^3 \left( \dfrac{x_\mathrm{max}^{2-\alpha}-x_\mathrm{min}^{2-\alpha}}{2-\alpha} \right) \equiv r_\sigma^3 g(\alpha,x_\mathrm{max},x_\mathrm{min}) \;.  \label{eq:g}
\end{align}
As $g(\alpha,r_\sigma,r_\mathrm{max},r_\mathrm{min})$ becomes singular for $\alpha=2$, we check that the limit of Equation~\ref{eq:g} exists. It exists and yields
\begin{equation}
g(2,x_\mathrm{max},x_\mathrm{min}) = \ln(x_\mathrm{max}) - \ln(x_\mathrm{min}) \;.
\end{equation}

\section{Derivation of the most likely $\alpha$, $r_\sigma$, $r_\mathrm{max}$, and $r_\mathrm{min}$}
\label{app:mle}

We abbreviate $x = x(r) = r/r_\sigma$ in Equation~\ref{eq:likelihood} 
\begin{equation}
\mathcal{L}(\alpha) = \ln \left( \prod \limits_{j=1}^{n_\mathrm{p}} p(r_j) \right) = \sum  \limits_{j=1}^{n_\mathrm{p}} \ln \left( N(\alpha) \, x_j^{-(\alpha+1)}  \right) \;,
\end{equation}
to obtain its derivative
\begin{align}
\dfrac{\mathrm{d} \mathcal{L}(\alpha)}{\mathrm{d} \alpha} &= \dfrac{\mathrm{d}}{\mathrm{d} \alpha} \sum  \limits_{j=1}^{n_\mathrm{p}}\left\{  \ln (N(\alpha)) + \ln \left( x_j^{-(\alpha+1)}  \right) \right\} \\
&= n_\mathrm{p} \dfrac{\mathrm{d}  \ln (N(\alpha))}{\mathrm{d} \alpha} -  \sum \limits_{j=1}^{n_\mathrm{p}} \ln \left( x_j  \right) \\
&= \dfrac{n_\mathrm{p}}{N(\alpha)} \dfrac{\mathrm{d}N(\alpha)}{\mathrm{d} \alpha} -  \sum \limits_{j=1}^{n_\mathrm{p}} \ln \left( x_j  \right) \;.
\end{align}
Inserting Equation~\ref{eq:N_sol} into the first term, we find
\begin{align}
\dfrac{\mathrm{d} \mathcal{L}(\alpha)}{\mathrm{d} \alpha} = - \dfrac{n_\mathrm{p}}{g(\alpha,x_\mathrm{max},x_\mathrm{min})} \dfrac{\mathrm{d}g(\alpha,x_\mathrm{max},x_\mathrm{min})}{\mathrm{d} \alpha} -  \sum \limits_{j=1}^{n_\mathrm{p}} \ln \left( x_j  \right) \;.
\label{eq:im1}
\end{align}

\com{In the same manner, we now derive Equation~\ref{eq:dLdrsigma}, starting from Equation~\ref{eq:likelihood}. 
We only write the variables of interest explicitly as function arguments for the sake of clarity, like $g(r_\sigma)$ instead of $g(\alpha, r_\sigma, r_\mathrm{max}, r_\mathrm{min})$. 
For the first term, depending on $N$, we need to calculate its derivative with respect to $r_\sigma$ to obtain
\begin{align}
\dfrac{\mathrm{d}N(r_\sigma)}{\mathrm{d} r_\sigma} &= \dfrac{\mathrm{d}}{\mathrm{d} r_\sigma} \left( (4\pi r_\sigma^3 g(r_\sigma))^{-1} \right) = - \dfrac{-3}{4\pi r_\sigma^4 g(r_\sigma)} + \dfrac{1}{4\pi r_\sigma^3} \left( - \dfrac{1}{g(r_\sigma)^{2}} \dfrac{\mathrm{d}g(r_\sigma)}{\mathrm{d} r_\sigma} \right) \\\
&= - N(r_\sigma) \left( \dfrac{3}{r_\sigma} + \dfrac{1}{g(r_\sigma)} \dfrac{\mathrm{d} g(r_\sigma)}{\mathrm{d} r_\sigma} \right) \;,
\label{eq:dN_rsigma}
\end{align}
in which we can calculate the last term using Equation~\ref{eq:f}
\begin{align}
\dfrac{\mathrm{d} g(r_\sigma)}{\mathrm{d} r_\sigma} =  \dfrac{\mathrm{d}}{\mathrm{d} r_\sigma} \left( \dfrac{r_\sigma^{\alpha-2}}{2-\alpha} \left( r_\mathrm{max}^{2-\alpha} - r_\mathrm{min}^{2-\alpha} \right) \right) = \dfrac{\alpha-2}{r_\sigma} g(r_\sigma)  \;.
\label{eq:dgg}
\end{align}
Inserting Equation~\ref{eq:dgg} into Equation~\ref{eq:dN_rsigma} we obtain
\begin{equation}
\dfrac{\mathrm{d}N(r_\sigma)}{\mathrm{d} r_\sigma} = - N(r_\sigma) \left( \dfrac{3}{r_\sigma} +  \dfrac{\alpha-2}{r_\sigma} \right)  =  - N(r_\sigma) \left( \dfrac{\alpha + 1}{r_\sigma} \right)
\end{equation}
Hence, the right-hand side of Equation~\ref{eq:dLdrsigma} is obtained when inserting this derivative into the first term of Equation~\ref{eq:dLdrsigma}.
}

\com{To obtain Equation~\ref{eq:dLdrmax}, we calculate
\begin{align}
\dfrac{1}{N(r_\mathrm{max})} \dfrac{\mathrm{d}N(r_\mathrm{max})}{\mathrm{d} r_\mathrm{max}} &= -\dfrac{1}{g(r_\mathrm{max})} \dfrac{\mathrm{d}g(r_\mathrm{max})}{\mathrm{d} r_\mathrm{max}} = -\dfrac{1}{g(r_\mathrm{max})} \dfrac{\mathrm{d}}{\mathrm{d} r_\mathrm{max}} \left( \dfrac{r_\sigma^{\alpha-2}}{2-\alpha} \left( r_\mathrm{max}^{2-\alpha} - r_\mathrm{min}^{2-\alpha} \right) \right) \\
&= - \dfrac{1}{g(r_\mathrm{max})} \left(\dfrac{r_\sigma^{\alpha-2}}{2-\alpha} \, (2-\alpha)\, r_\mathrm{max}^{1-\alpha} \right) = -(2-\alpha) \dfrac{r_\mathrm{max}^{1-\alpha}}{r_\mathrm{max}^{2-\alpha} - r_\mathrm{min}^{2-\alpha}} \;. \label{eq:res_rmax}
\end{align}
Analogously, we can derive the likelihood with respect to $r_\mathrm{min}$
\begin{align}
\dfrac{1}{N(r_\mathrm{min})} \dfrac{\mathrm{d}N(r_\mathrm{min})}{\mathrm{d} r_\mathrm{min}} &= -\dfrac{1}{g(r_\mathrm{min})} \dfrac{\mathrm{d}g(r_\mathrm{min})}{\mathrm{d} r_\mathrm{min}} = -\dfrac{1}{g(r_\mathrm{min})} \dfrac{\mathrm{d}}{\mathrm{d} r_\mathrm{min}} \left( \dfrac{r_\sigma^{\alpha-2}}{2-\alpha} \left( r_\mathrm{max}^{2-\alpha} - r_\mathrm{min}^{2-\alpha} \right) \right) \\
&= \dfrac{1}{g(r_\mathrm{min})} \left(\dfrac{r_\sigma^{\alpha-2}}{2-\alpha} \, (2-\alpha)\, r_\mathrm{min}^{1-\alpha} \right) = (2-\alpha) \dfrac{r_\mathrm{min}^{1-\alpha}}{r_\mathrm{max}^{2-\alpha} - r_\mathrm{min}^{2-\alpha}}  \;. \label{eq:res_rmin}
\end{align}
Since the resulting right-hand sides of Equations~\ref{eq:res_rmax} and \ref{eq:res_rmin} are equivalent after interchanging $r_\mathrm{max}$ and $r_\mathrm{min}$, their limiting behaviour is also the same, hence, there is also no preferred $r_\mathrm{min}$, as expected by construction of the approach.}

\section{Application of Stirling's formula for Equation~\ref{eq:alpha_fluid}}
\label{app:Stirling}

The sum in the denominator of Equation~\ref{eq:alpha_fluid} can be approximated as
\begin{align}
\sum \limits_{j=1}^{n_\mathrm{p}}  \ln \left( \dfrac{r_j}{r_\mathrm{max}} \right) \approx  \sum \limits_{j=1}^{n_\mathrm{p}}  \ln \left( \dfrac{j}{n_\mathrm{p}} \right) 
\end{align}
using Equation~\ref{eq:hom_n}.
This allows us to employ Stirling's approximation for $\ln(n_\mathrm{p}!)$ for large $n_\mathrm{p}$ as follows
\begin{align}
\sum \limits_{j=1}^{n_\mathrm{p}}  \ln \left( \dfrac{j}{n_\mathrm{p}} \right)& = \sum \limits_{j=1}^{n_\mathrm{p}}  \ln \left( j \right) - n_\mathrm{p} \ln (n_\mathrm{p}) = \ln(n_\mathrm{p}!) - n_\mathrm{p} \ln (n_\mathrm{p}) \\
&\approx n_\mathrm{p} \ln (n_\mathrm{p}) - n_\mathrm{p} + \mathcal{O}(\ln(n_\mathrm{p})) - n_\mathrm{p} \ln(n_\mathrm{p}) \\
&= - n_\mathrm{p} + \mathcal{O}(\ln(n_\mathrm{p})) \;.
\end{align}
Thus, Equation~\ref{eq:alpha_fluid} is approximated by
\begin{equation}
\alpha (n_\mathrm{p}) = 2 - \dfrac{n_\mathrm{p}}{n_\mathrm{p} - \mathcal{O}(\ln(n_\mathrm{p}))} \;.
\end{equation}

\end{document}